\documentclass{emulateapj}

\begin{document}

\newcommand{\cl}{{RDCS J1252.9-2927}}
\newcommand{\clt}{{RX J0152.7-1357}}
\newcommand{\ms}{{MS 1358+62}}
\newcommand{\etal}{{\em et~al.\,}}
\title{Evolution in the Cluster Early-type Galaxy Size-Surface Brightness Relation at $z \simeq 1$ \altaffilmark{1}}

\author{B. P. Holden\altaffilmark{2}}

\author{J. P. Blakeslee\altaffilmark{3}}

\author{M. Postman\altaffilmark{4}}

\author{G. D. Illingworth\altaffilmark{2}}

\author{R. Demarco\altaffilmark{3}}

\author{M. Franx\altaffilmark{5}}

\author{P. Rosati\altaffilmark{6}}

\author{R. J. Bouwens\altaffilmark{2}}

\author{A. R. Martel\altaffilmark{3}}

\author{H. Ford\altaffilmark{3}}

\author{M. Clampin\altaffilmark{7}}

\author{G.F. Hartig\altaffilmark{4}}

\author{N. Ben\'{i}tez\altaffilmark{8}}

\author{N. J. G. Cross\altaffilmark{3}}

\author{N. Homeier\altaffilmark{3}}

\author{C. Lidman\altaffilmark{9}}

\author{F. Menanteau\altaffilmark{3}}

\author{A. Zirm\altaffilmark{5}}

 \author{D.R. Ardila\altaffilmark{3}}
 \author{F. Bartko\altaffilmark{10} }
 \author{L.D. Bradley\altaffilmark{3}}
 \author{T.J. Broadhurst\altaffilmark{11}}
 \author{R.A. Brown\altaffilmark{4}}
 \author{C.J. Burrows\altaffilmark{4}}
 \author{E.S. Cheng\altaffilmark{12}}
 \author{P.D. Feldman\altaffilmark{3}}
 \author{D.A. Golimowski\altaffilmark{3}}
 \author{T. Goto\altaffilmark{3}}
 \author{C. Gronwall\altaffilmark{13}}
 \author{L. Infante\altaffilmark{14}}
 \author{R.A. Kimble\altaffilmark{7}}
 \author{J.E. Krist\altaffilmark{4}}
 \author{M.P. Lesser\altaffilmark{15}}
 \author{D. Magee\altaffilmark{2}}
 \author{S. Mei\altaffilmark{3}}
 \author{G.R. Meurer\altaffilmark{3}}
 \author{G.K. Miley\altaffilmark{5}}
 \author{V. Motta\altaffilmark{14}}
 \author{M. Sirianni\altaffilmark{4}}
 \author{W.B. Sparks\altaffilmark{4}}
 \author{H.D. Tran\altaffilmark{16}}
 \author{Z.I. Tsvetanov\altaffilmark{3}}
 \author{R.L. White\altaffilmark{4}}
 \author{W. Zheng\altaffilmark{3}}

\altaffiltext{1}{Based on observations with the NASA/ESA Hubble Space
Telescope, obtained at the Space Telescope Science Institute, which is
operated by the Association of Universities for Research in Astronomy,
Inc. under NASA contract No. NAS5-26555.  Based on observations
obtained at the European Southern Observatory using the ESO Very Large
Telescope on Cerro Paranal (ESO Large Programme 166.A-0701).}

\altaffiltext{2}{Department of Astronomy, UCO/Lick Observatories,
University of California, Santa Cruz, 95064; holden@ucolick.org}
 \altaffiltext{3}{Department of Physics and Astronomy, Johns Hopkins
 University, 3400 North Charles Street, Baltimore, MD 21218.}
 \altaffiltext{4}{STScI, 3700 San Martin Drive, Baltimore, MD 21218.}
 \altaffiltext{5}{Leiden Observatory, Postbus 9513, 2300 RA Leiden,
 Netherlands.}
 \altaffiltext{6}{European Southern Observatory,
 Karl-Schwarzschild-Strasse 2, D-85748 Garching, Germany.}
 \altaffiltext{7}{NASA Goddard Space Flight Center, Code 681,
   Greenbelt, MD 20771.} 
 \altaffiltext{8}{Inst\'{i}tuto de Astrof\'{i}sica de Andaluc\'{i}a,
   Camino Bajo de Hu\'{e}tor 24, Granada, 18008, Spain.} 
 \altaffiltext{9}{European Southern Observatory,
 Alonso de Cordova 3107, Casilla 19001, Santiago, Chile.}
 \altaffiltext{10}{Bartko Science \& Technology, 14520 Akron Street, 
 Brighton, CO 80602.}	
 \altaffiltext{11}{Racah Institute of Physics, The Hebrew University,
 Jerusalem, Israel 91904.}
 \altaffiltext{12}{Conceptual Analytics, LLC, 8209 Woburn Abbey Road, Glenn Dale, MD 20769}
 \altaffiltext{13}{Department of Astronomy and Astrophysics, The
 Pennsylvania State University, 525 Davey Lab, University Park, PA
 16802.}
 \altaffiltext{14}{Departmento de Astronom\'{\i}a y Astrof\'{\i}sica,
 Pontificia Universidad Cat\'{\o}lica de Chile, Casilla 306, Santiago
 22, Chile.}
 \altaffiltext{15}{Steward Observatory, University of Arizona, Tucson,
 AZ 85721.}
 \altaffiltext{16}{W. M. Keck Observatory, 65-1120 Mamalahoa Hwy., 
 Kamuela, HI 96743}

\begin{abstract}

We investigate the evolution in the distribution of surface
brightness, as a function of size, for  elliptical and S0
galaxies in the two clusters \cl, $z=1.237$ and \clt, $z=0.837$.  We
use multi-color imaging with the Advanced Camera for Surveys on the
Hubble Space Telescope to determine these sizes and surface
brightnesses.  Using three different estimates of the surface
brightnesses, we find that we reliably estimate the surface brightness
for the galaxies in our sample with a scatter of $< 0.2$ mag and with
systematic shifts of $\lesssim 0.05$ mag.  We construct samples of
galaxies with early-type morphologies in both clusters.  For each
cluster, we use a magnitude limit in a band which closely corresponds
to the rest-frame $B$, to magnitude limit of $M_B = -18.8$ at $z=0$,
and select only those galaxies within the color-magnitude sequence of
the cluster or by using our spectroscopic redshifts.  We measure
evolution in the rest-frame $B$ surface brightness, and find $-1.41
\pm 0.14$ mag from the Coma cluster of galaxies for \cl\ and $-0.90
\pm 0.12$ mag of evolution for \clt, or an average evolution of
$(-1.13 \pm 0.15) z$ mag.  Our statistical errors are dominated by the
observed scatter in the size-surface brightness relation, $\sigma =
0.42 \pm 0.05$ mag for \clt\ and $\sigma = 0.76 \pm 0.10$ mag for \cl.
We find no statistically significant evolution in this scatter, though
an increase in the scatter could be expected.  Overall, the pace of
luminosity evolution we measure agrees with that of the Fundamental
Plane of early-type galaxies, implying that the majority of massive
early-type galaxies observed at $z \simeq 1$ formed at high redshifts.

\end{abstract}

\section{Introduction}

The apparent evolution in the colors and magnitudes of early-type
cluster galaxies has long been used as a test of the history of galaxy
evolution.  The relations between magnitude and size or velocity
dispersion play an important role, as they give the observer a way to
predict the apparent luminosity of a galaxy based on some other
observable property \citep{faber76,tully77,kormendy77,djorgovski87}.
Using these scaling relations, it has been observed that there is
remarkable uniformity in cluster early-type galaxy properties at low
redshifts, $z \la 0.2$, \citep[for
example]{sandage91,jfk96,bernardi2003}.  At redshifts up to $\simeq
1$, there have been a number of papers that have established a
decrease in the mass-to-light ratio, using the Fundamental Plane.
These studies find that the trend with redshift corresponds to a
passively evolving stellar population which has a luminosity-weighted
last epoch of major star-formation around $z \simeq 2-3$
\citep{vandokkum96,kelson97,vandokkum1998,kelson2000c,pvd_mf2001,vandokkum2003,wuyts2004,holden2005}.
These systems also appear to have old stellar populations when
examined using their colors \citep[{\em
e.~g.};][]{bower92b,aragon93,ellis97,stanford98,blakeslee2003,holden2004},
with a similar last epoch of star formation of $z \simeq 2-3$.
Combining the evolution in colors and in the scaling relations provide
us with the data to measure the luminosity-weighted age of early-type
galaxies and, therefore, the epoch when most of the constituent stars
formed.

The size-magnitude relation is a projection of the Fundamental Plane
(FP) along the velocity dispersion axis.  It has long been known that
low redshift early-type galaxies follow such a relation
\citep{kormendy77}.  By examining how the apparent magnitude changes
as a function of redshift at a fixed size, we can measure the apparent
luminosity evolution in early-type galaxies.  Earlier measurements of
this at $z\simeq 0.4-0.6$ \citep{sandage90a,sandage90b,sandage91}, and
at $z\simeq 0.7-1.0$
\citep{schade96,schade97,schade99,sandage2001,lubin2001a,lubin2001b,lubin2001c},
found the same pace of evolution as found by the evolution of colors
or from the FP of early-type galaxies.

We investigate the size-surface brightness relation for two
complementary goals.  First, we can extend the measured evolution to
$z=1.237$, which corresponds to the highest redshifts that the FP has
been measured to date \citep{vandokkum2003,holden2005}.  However,
because we have a much larger sample, we can also examine the
distribution of the size-surface brightness relation to see if we find
evidence for early-type galaxies that have younger stellar
populations.  Traditionally, this has been done with colors or
spectroscopically \citep[see ][for ``E+A'' galaxies that have
early-type morphologies]{tran2003}.  \citet{vandokkum2003}, however,
find that one of their early-type galaxies has an ``E+A'' like
spectrum and appears to be an outlier from the distribution of sizes
and magnitudes for that cluster \citep{holden2004}.  A significant
population of such objects would imply that we are finding the epoch
when intermediate mass cluster galaxies are completing their evolution
into cluster early-type galaxies.

Our sample has two clusters, \cl\ at $z=1.237$
\citep{rosati2004} and \clt\ at $z=0.837$ \citep{dellaCeca00}.  For
each we use Hubble Space Telescope imaging data and other supporting
ground-based data, discussed in \S 2, to construct a sample of
early-type cluster galaxies.  For each galaxy in our sample, we
measure three different total magnitudes and two different half-light
radii ($r_h$).  The comparison of these measurements are discussed in
\S 3.  We verify that our sizes and total magnitudes are consistent,
then use those two quantities to measure the mean surface brightness
interior to the galaxy's half-light radius ($<\mu_h>$).  In \S 4, we
use $r_h$ and $<\mu_h>$ to estimate the amount of luminosity evolution
seen from low redshift to the clusters in our sample.  In addition, we
examine the scatter in the relation and search for galaxies out of the
norm.  We discuss and summarize, in \S 5, our results in the context
of other work.  Throughout the paper we compute distances and absolute
magnitudes using the cosmological parameters of $\Omega_m = 0.3$,
$\Omega_{\Lambda} = 0.7$ and ${\rm H_o = 70\ km\ s^{-1}\ Mpc^{-1}}$.
We use the AB magnitude system \citep{oke90} unless otherwise noted.

\section{Data}

We combine a number of different datasets to explore the evolution of
the size-magnitude relation.  For our high redshift samples, we use
the Advanced Camera for Surveys (ACS) imaging data for \cl, already
discussed in \citet{blakeslee2003}, and for \clt.  Both clusters are
described in \citet{postman2004} which focuses specifically on
morphological classification and the morphology-density relation.
Both clusters have additional ground-based data in the form of spectra
and multi-wavelength imaging.  We compare our high redshift result
with the data for \ms, at $z=0.328$, from \citet{kelson2000a}.  We
select this particular dataset because of the similarities in methods
used by the authors.

\subsection{RDCS J1252.9-2927}
\label{acs}

\begin{figure}[tbp]
\begin{center}
\includegraphics[width=3.0in]{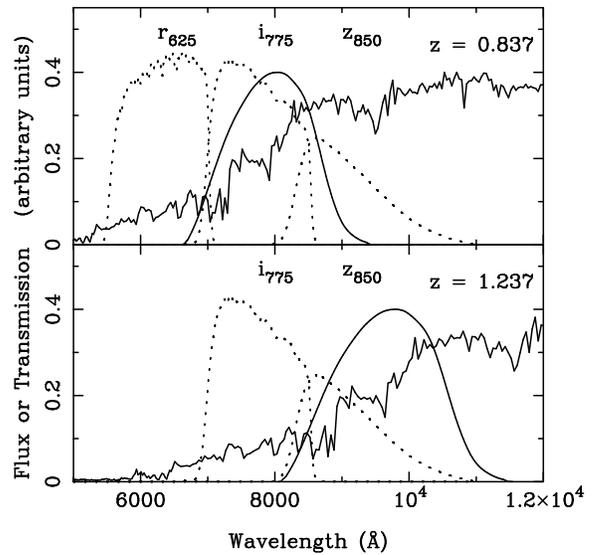}
\end{center}
\caption[f1.eps]{A 13 Gyrs old early-type galaxy spectral
  energy distribution at a redshift of $z=0.837$, top, and $z=1.237$,
  bottom.  In both panels, we plot the filters we used for the
  observations with dotted lines and the Johnson $B$ filter as a solid
  line in the rest-frame of the galaxy.  In each case, we have a color
  that straddles the 4000 \AA\ break and a filter that overlaps with
  the Johnson $B$.}
\label{sed}
\end{figure}

As described in \citet{blakeslee2003}, the Advanced Camera for Surveys
(ACS) imaged \cl, at $z=1.237$, for 12 orbits in the F775W filter, or
$i_{775}$, and 20 orbits in the F850LP filter, or $z_{850}$.  The
cluster was imaged as a mosaic of two by two positions, with three
orbits in $i_{775}$ and five orbits in $z_{850}$ at each position.
The central area was covered by each of the four positions, resulting
in deeper imaging over the central arcminute.  In Figure \ref{sed}, we
plot the filter transmission curves and a representative spectral
energy distribution.  We note here that the 850 in the F850LP
designation refers to the wavelength where the filter begins.
The central wavelength of the filter, when combined with the CCD
performance is 9077\AA, sampling at 4058\AA\ in the rest-frame of \cl.

The data were processed using APSIS, the ACS pipeline science
investigation software, described in \citet{blakeslee_pipe2003}.
This pipeline generates drizzled images in each band, a detection
image made by combining all of the drizzled data, and photometric
catalogs constructed using a combination of the detection image and
the final images in each filter.  

As discussed in \citet{blakeslee2003}, we selected objects in \cl\
with $z_{850} < 24.8$ to ensure adequate signal-to-noise for the fits.
Each galaxy was fit with a single component S{\'e}rsic model
\citep{sersic} using v1.7a, Rev 3 of the {\tt GALFIT} package
\citep{peng2002}.  The approach of \citet{peng2002} is to construct a
model, convolve it with a point spread function and then compare the
$\chi^2$ of the resulting model with the data.  For a noise model, we
use the variance maps produced by the APSIS pipeline.  Our
point spread functions are based on ACS imaging data of crowded star
fields.  We will discuss this further in \S \ref{model}.

Each galaxy with $0.5 < i_{775}-z_{850} < 1.2$ was morphologically
identified by four of us (BH, NC, MP, and MF).  These identifications
will be discussed in another paper \citep{postman2004}.  The
identifications were made using the T-type morphological
classification system \citep{devauc1991}.  For the classifications, we
found that 75\% of the time all four classifiers agreed on the
morphological type, and three out of four agreed 80\% of the time.
These statistics are for the whole of the sample and the agreement is
much higher for the brighter galaxies.  However, what matters for this
work is how robust the classification for early-type galaxies in the
final sample.  There are only two galaxies in \cl\ that appear in our
final sample where a majority of the four classifiers did not agree on
a classification as an early type ({\em i.e.}; lenticular or
elliptical).  Both of those galaxies appear near the magnitude limit
of our sample and neither are part of the spectroscopic sample we will
discuss below.  At $z_{850} < 23$, 15 out of the 20 early-type
galaxies were classified unanimously.  From $24.5 > z_{850} > 23$, the
fraction drops to 20 unanimous identifications out of 32 early-type
galaxies, or 63\%, but for 30 out of 32 times at least three out of
the four classifiers agreed, higher than the 80\% average for the
whole sample.

Despite the high degree of internal consistency, one potential concern
is a systematic offset in our classifications as compared with other
publications.  To control for this, one of the classifiers, MP,
assigned types to every galaxy in the \ms\ sample of
\citet{fabricant2000}.  The classifications by MP of \ms\ agreed 80\%
of the time with those from \citet{fabricant2000}.

\begin{figure}[tbp]
\begin{center}
\includegraphics[width=3.0in]{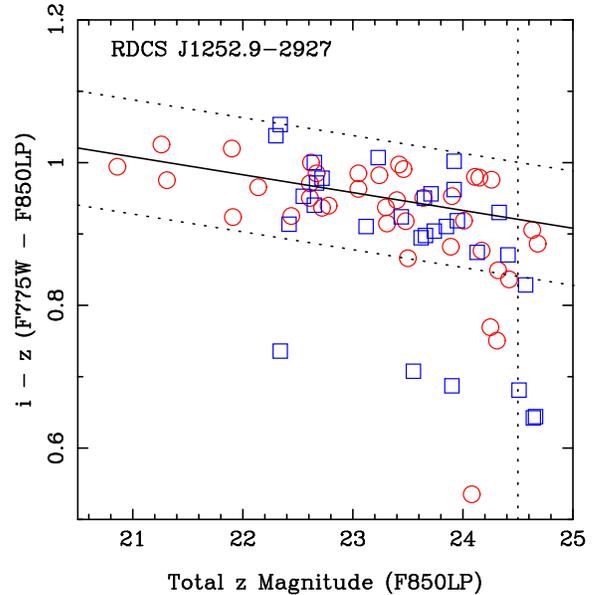}
\end{center}
\caption[f2.eps]{Color-magnitude diagram for the early-type galaxies
  in the ACS imaging data for \cl\ \citep{blakeslee2003}. The colors
  are measured within the half-light radius as measured from fitting
  S\'{e}rsic models.  The red circles represent elliptical galaxies,
  while the blue squares are the galaxies classified as S0's.  The
  solid line is the mean color-magnitude relation while the parallel
  dotted lines represent the $2\sigma$ limits for the color-selected
  sample.  The vertical dotted line is the magnitude limit for the
  color-selected sample.  }
\label{clcm}
\end{figure}

In Figure \ref{clcm}, we plot the color-magnitude diagram of the
early-type galaxies with $T < 0$ that are within 1 Mpc, or 1\farcm 92,
of the cluster center and have $z_{850} < 24.5$ mag.  The circles
 represent galaxies classified as ellipticals ($-5 \le T \le
-4$) and squares represent S0 ($-3 \le T \le 1$) galaxies.  The solid
line is the color-magnitude relation $(i-z) = 0.958 - 0.025 (z_{850} -
23)$.  The dotted diagonal line represents two standard deviations, or
0.08 mag, from this relation, see \citet{blakeslee2003} for details.
We will use this color criteria for selecting early-type galaxies in
the cluster.  We remove spectroscopic non-members that, nonetheless,
have the colors of the early-type red sequence in \cl.  We plot the
magnitude limit for morphological identification, $z_{850} = 24.5$, as
a vertical, dotted line in Figure \ref{clcm}. This magnitude limit
corresponds to $\simeq 0.2 L^{\star}$.  We calculate $L^{\star}$
starting with the value from \citet{norberg2002} and evolving it with
the relation of \citet{vandokkum2003}.  At $z=1.24$, $M^{\star}_{B} =
-21.7$ in AB magnitudes using the previous results
\citep{blakeslee2003}.

In addition to the ACS imaging data described above, we also use the
photometric and spectroscopic data collected in
\citet{demarco2003PhDT,rosati2004,lidman2004,toft2004}.  The catalog
for \cl\ contains photometry in the B, V, R, i, z, J and ${\rm K_{s}}$
bands.  The near-infrared images were obtained with ISAAC on the VLT
\citep{lidman2004}, while the optical images were observed using the
FORS1 and FORS2 \citep{fors} instrument on the VLT.  These optical and
near-infrared images were used to create a sample of galaxies for a
redshift survey of potential cluster members.  All galaxies have ${\rm
K_{s}} < 21$ and $J - K < 2.1$ and $R - K > 3$
\citep{demarco2003PhDT,rosati2005}.  These colors were chosen to
preferentially select galaxies at the redshift of the cluster
regardless of the spectral energy distribution while ignoring redder
stars as well as low redshift systems.  A total of 383 objects were
observed, of which 235 have secure redshifts.  This yields a sample of
36 cluster members.  The sample is almost complete for $z_{850} < 23$,
which is $\simeq 0.8 L^{\star}$.  When comparing other samples with
the redshift sample of \cl, we will move the magnitude limit to
$z_{850} < 23$ but we will not use the color selection, shown with the
two dotted lines, of Figure \ref{clcm}.  This only adds one object to
the sample that would not be included in the color-selected sample.

\subsection{RX J0152.7-1357}

\begin{figure*}[tbp]
\begin{center}
\includegraphics[width=6.0in]{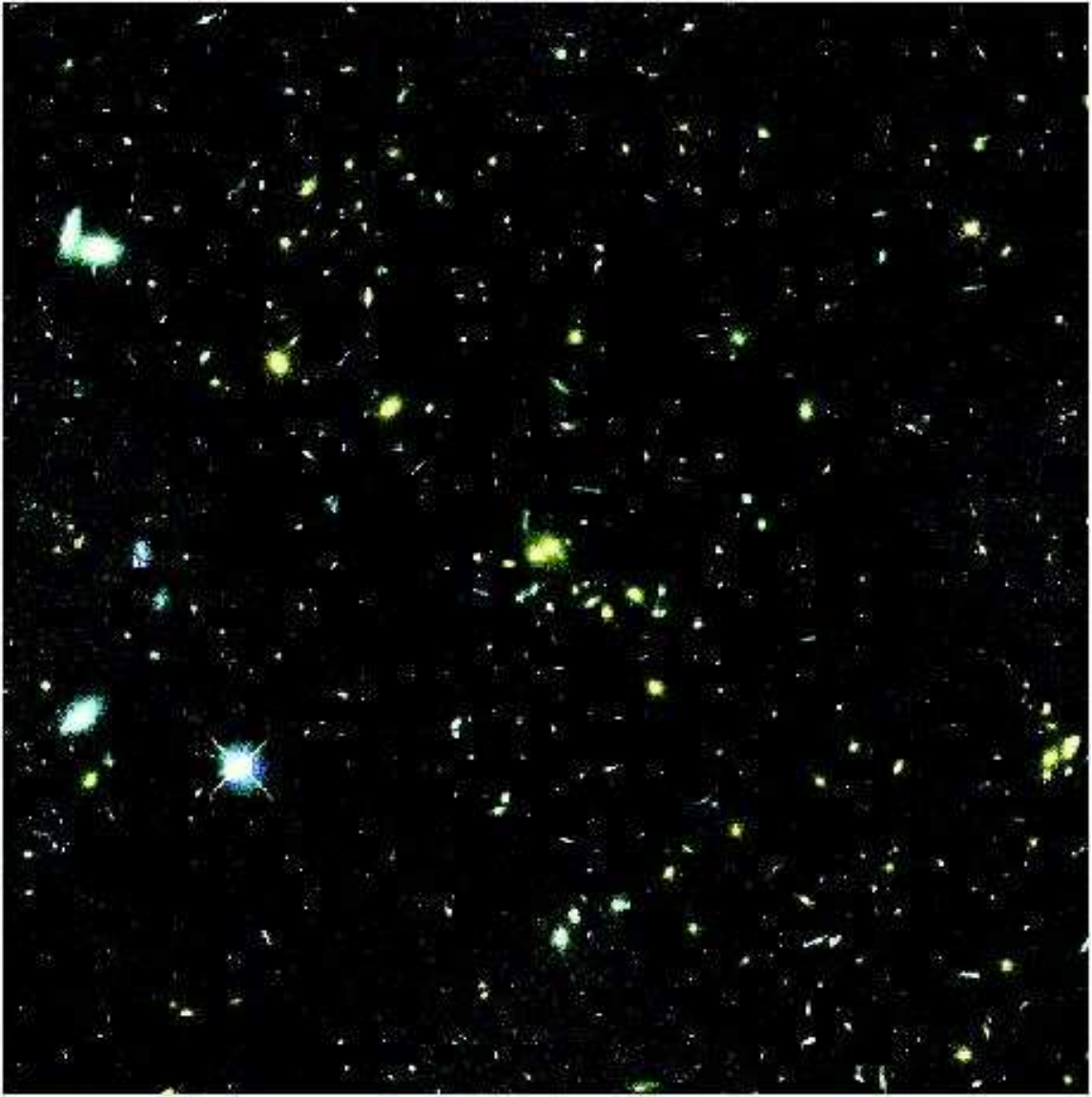}
\end{center}
\caption[my_cl0152_acs.eps]{Color image made from combining the $r_{625}$,
  $i_{775}$ and $z_{850}$ imaging data for \clt.  The image is
  2\arcmin on a side, with North up and East to the left.  Some 
  ``red'' galaxies in this image are actually members of a foreground
  group at $z=0.64$. }
\label{cltmos}
\end{figure*}

For \clt , at $z=0.837$, the ACS images were taken
with the F625W, or $r_{625}$, filter along with the $i_{775}$ and
$z_{850}$ filters.  Each of the three filters was observed for two
orbits.  As with \cl, \clt\ was observed using a two-by-two mosaic
pattern with deeper imaging in the cluster core.  In Figure
\ref{cltmos}, we show the inner 2\arcmin\ around the cluster center.
This color image was made from the combination of all three filters.  

\begin{figure}[tbp]
\begin{center}
\includegraphics[width=3.0in]{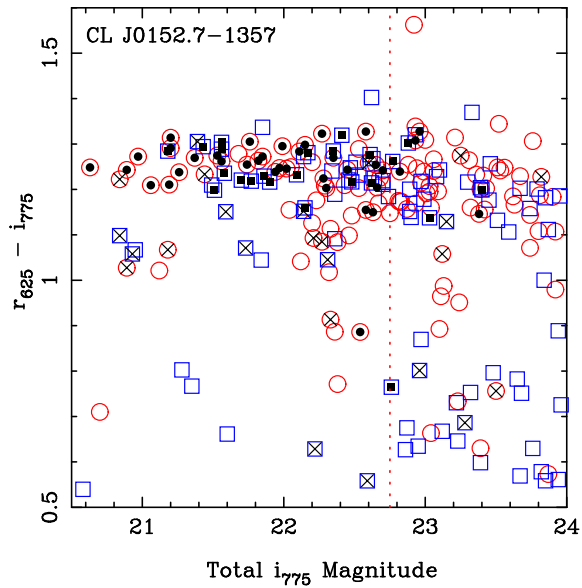}
\end{center}
\caption[f3.eps]{Color-magnitude diagram for the early-type
  galaxies in the ACS imaging data for \clt.  The colors are measured
  within the half-light radius.  Red circles represent elliptical galaxies
  while blue squares are the galaxies classified as S0. The dotted line
  represents the magnitude limit.  Symbols with solid dots are
  confirmed to be cluster members while points with crosses are
  non-members.  Unlike \cl, we did not select galaxies using colors
  because of the presence of a $z=0.64$ group of galaxies.  Known
  members of this group can be seen as crosses with colors very close
  to the ``red-sequence'' of the \clt\ members.  }
\label{cltcm}
\end{figure}

Our sample for \clt\ was constructed in as similar a manner as
possible as our sample for \cl.  We used the SExtractor detection
catalog to find galaxies.  All objects with isophotal $i_{775} < 24.8$
were fit with a S{\'e}rsic model with $1 \le n \le 4$ and visually
typed \citep{postman2004}.  The model fits were determined in the
$i_{775}$ filter as that filter corresponds closest to the rest-frame
$B$ (see Figure \ref{sed}), and the resulting apertures were used to
measure the apparent colors.  These results are shown in Figure
\ref{cltcm}.

For \clt, we have obtained B, V, R, and I imaging data using the Keck
telescope with the LRIS \citep{oke95}.  We used SOFI \citep{sofi} on
the New Technology Telescope \citep{demarco2005a} to observe in the J
and ${\rm K_{s}}$ bands.  Using FORS1 and FORS2 on the VLT,
\citet{demarco2005a} targeted galaxies with $R \le 24$ using the
photometric redshift selection criterion of $0.7 \le z_{phot} \le
0.95$.  262 objects were observed resulting in 102 cluster members out
of 227 measured redshifts.

As discussed in \citet{postman2004}, the classifications for \clt\
were done by three of us (MP, NC, and BH).  The error rates are
effectively the same as for \cl\ but we examined all galaxies down to
$i_{775} < 24$, a fainter limit than our completeness for the
redshift survey (discussed below).  

In the field of \clt , \citet{demarco2005a} found a group at $z=0.64$
which has a significant number of early-type galaxies.  These objects
are, at best, slightly bluer than the cluster members (see Figure
\ref{cltcm}).  The inclusion of these early-type galaxies in the
photometric redshift selected sample reaffirms this.  As such, we will
not make a color selected sample for \clt\ but, rather, will only use
the spectroscopic redshift sample.  In the magnitude range of $20 < i_{775} <
24$, the completeness of early-type galaxies ranges from 80\% to 0\%.
The effective magnitude limit of the survey is $i_{775} < 23$ as there
is only one early-type galaxy fainter than that with a measured
redshift.  In the range $20.25 < i_{775} < 22.75$, roughly 70\% of the
early-type galaxies with spectra are cluster members.  In that same
magnitude range, 60\% of all early-type galaxies have spectra.  This
completeness is roughly the same as \ms\ \citep{kelson2000c} against
which we will compare our data.  However, it should be realized that
the photometric redshift selection imposes, in effect, a color
selection as photometric redshifts are based on colors.  Objects with
stronger 4000 \AA\ breaks are more likely to be selected for
spectroscopic follow-up.  For the remainder of the paper, we select
all early-type galaxies in \clt\ with $i_{775} < 22.75$ that are
spectroscopically confirmed cluster members.  This corresponds to
$\simeq 0.5 L^{\star}$.

\subsection{MS 1358+62}

For our sample in \cl\ and \clt, we used filters that
straddled the 4000 \AA\ break for cluster members (see Figure
\ref{sed}). To measure the amount of apparent luminosity evolution, at
a fixed size, requires that we measure the magnitudes and sizes in
filters that match, as closely as possible, the same rest-frame band pass.

For a low redshift sample, we use the results from MS1358+62, at
$z=0.33$, from \citet{kelson2000a,kelson2000b,kelson2000c} because of
the thorough study of the surface brightness profiles in the
rest-frame $B$ band.  For \ms , \citet{kelson2000a,kelson2000b}
conducted a spectroscopic survey to determine the velocity dispersion
for a flux-limited sample of galaxies.  The 52 objects were selected
from the \citet{fisher1998} study of \ms\ to $R\leq 21$.  The total
sample of \citet{kelson2000a,kelson2000b} includes an additional
three galaxies that are below the magnitude limit, but we used only
those with $R\leq 21$.  The selection for this sample was done
regardless of morphology, so we include only galaxies that are
classified earlier than 0, or the S0/a morphological class, according
to the scheme of \citet{devauc1991}.

We use the tabulated values of the observed rest-frame $V$ band
surface brightnesses from \citet{kelson2000a} along with the
half-light radii in that paper.  To convert the $V$ band surface
brightness into a $B$ band surface brightness, we use the color from
\citet{kelson2000c} \citep[taken from][]{vandokkum1998}.  The
apertures used to define these colors are the half-light radii
measured by fitting a de Vaucouleurs profile.  In addition, the
authors fit S{\'e}rsic models.  There is a difference in their
S{\'e}rsic models, however, when compared to the S{\'e}rsic models in
this paper.  In \citet{kelson2000a} the authors allowed the S{\'e}rsic
parameter to the range $1 \le n \le 6$.  We restricted $n$ over a
smaller range, $1 \le n \le 4$ for our fits of \cl\ and \clt.  So, in
all cases where \citet{kelson2000a} finds $n > 4$, we replace the
results of that fit, namely $r_h$ and $<\mu_h>$, with corresponding
$n=4$ results from \citet{kelson2000a}.

\section{Measures of the size and magnitude}

In order to measure the evolution in the Kormendy relation
\citep{kormendy77}, we need robust measures of the half-light radius
and surface brightness.  Below we discuss two methods, one parametric
and one non-parametric, we used to measure both the total magnitudes,
from which we infer the surface brightness, and the half-light radii.

\subsection{Model Fitting}
\label{model}

To determine the colors of each of the galaxies, we used models to
establish the appropriate size of the apertures.  We fit a single
S{\'e}rsic profile and allowed the parameter $n$ to vary over the
range $1 \le n \le 4$.  We imposed this restriction because we found
that, if we did not, a few objects would settle to very high $n$
numbers and, correspondingly large half-light radii.  As the goal for
the fits was to provide radii for measuring the colors of the
red-sequence galaxies, large radii potentially could lead to
systematic errors in the measurements.

The models were fitted using {\tt GALFIT} \citep{peng2002}, and yield
the half-light radius and total magnitude of the model.  The mean
interior surface brightness within the half-light radius for that
model can then be computed as \( <\mu_{h}> = m + 5 \log_{10} r_{h} +
2.5 \log_{10} 2\pi \) \citep{jfk95sb}.  The {\tt GALFIT} output for
the size is actually the semi-major axis of the best fitting
elliptical model.  We convert this radius into the effective radius,
$r_{h} = \sqrt{ab}$, where $a$ and $b$ are the semi-major and
semi-minor axes respectively, when we discuss the half-light radius.
As mentioned above, the fitting procedure constructs a model,
convolves it with the point spread function, and then compares the
resulting $\chi^2$ between the model and the data.  Because the
modeling process explicitly includes the point spread function, even
for objects with sizes smaller than the point spread function, a full
width at half maximum of $\simeq$ 0\farcs09, could be, in principle,
reliably measured.  However, errors in our point spread functions
could cause systematic biases in our resulting half-light radii.
Thus, we will exclude from our analysis objects with model half-light
radii less than 0\farcs 1 for the rest of the paper.

One of the advantages of the {\tt GALFIT} software is the ability to
fit multiple component models to multiple galaxies simultaneously.
For most of the galaxies, however, we masked out neighboring objects
using the results of the SExtractor segmentation map, {\em i.~e.}, the
map of pixels above the isophotal detection threshold.  At minimum,
each postage stamp was 50 by 50 pixels.  The actual size was four
times the square root of SExtractor's isophotal area, the number of
pixels one $\sigma$ above the sky.  For close galaxies, however, we
had to fit models simultaneously.  This was often the case for the
luminous ellipticals in the cores of the clusters.  In cases where a
multiple galaxies had to be fit, we ensured that all of the pixels
that would be in the individual stamps using the above recipe were
included in the stamp containing multiple galaxies.

\begin{figure}[tbp]
\begin{center}
\includegraphics[width=3.0in]{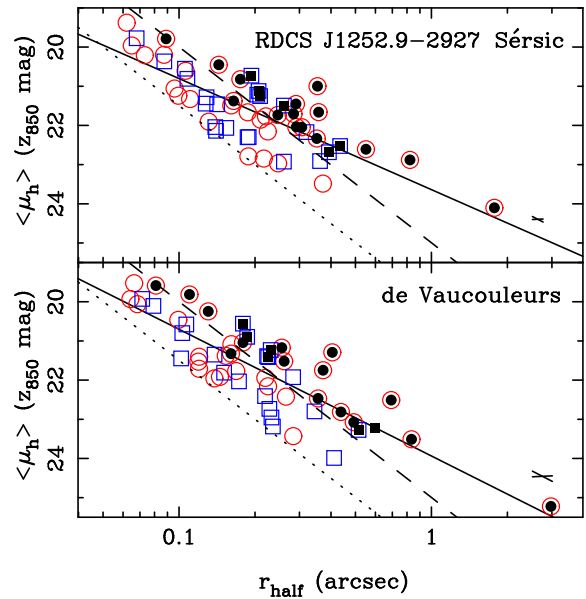}
\end{center}
\caption[f4.eps]{Size-surface brightness relation for the early-type
  galaxies in the ACS imaging data of \cl.  Filled symbols are
  spectroscopically selected members and the open symbols are color
  selected, red circles for ellipticals and blue squares for S0's.
  The surface brightnesses are in $z_{850}$ magnitudes.  We show the
  results for S{\'e}rsic models with $1 \le n \le 4$ in the top
  diagram while the bottom diagram shows the sizes and magnitudes
  estimated using de Vaucouleurs models.  The best fitting relation to
  the whole sample is shown as a solid line, while the magnitude limit
  of the color-selected sample is shown as a dotted line and the
  magnitude limit we impose for the spectroscopically selected sample
  is shown with a dashed line, see \S \ref{acs} for details. We plot
  the semi-major axis of the ellipse of the anti-correlation between
  $r_h$ and $<\mu_h>$ as a line.  The slope of the anti-correlation is
  very close to the slope of the size-surface brightness relation.  We
  do note that the sample appears incomplete at the smallest sizes,
  $r_h <$0\farcs1.  }
\label{clsizemag}
\end{figure}

\begin{figure}[tbp]
\begin{center}
\includegraphics[width=3.0in]{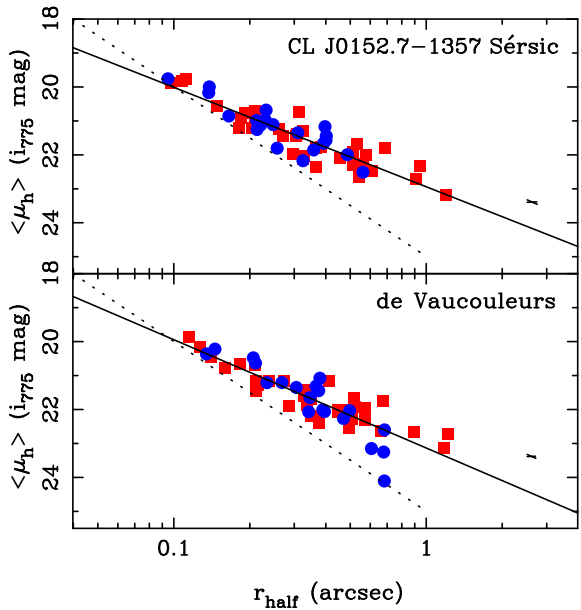}
\end{center}
\caption[f5.eps]{Size-surface brightness relation for the early-type
  galaxies in the ACS imaging data of \clt, shown with filled circles.
  As with Figure \ref{clsizemag}, red circles are ellipticals and blue
  squares are S0's.  The surface brightnesses are in $i_{775}$
  magnitudes. We plot the results for S{\'e}rsic models with $1 \le n
  \le 4$ in the top diagram while the bottom diagram shows the sizes
  and magnitudes estimated using de Vaucouleurs models.  The solid
  line shows the mean relation while the dotted shows the magnitude
  limit.  As with Figure \ref{cltsizemag}, we plot the semi-major axis
  of the ellipse of the anti-correlation between $r_h$ and $<\mu_h>$
  as a line.  \clt appears less incomplete at small sizes, possibly
  because of the brighter magnitude limit imposed by the redshift
  survey used for selection.}
\label{cltsizemag}
\end{figure}

For every object in the color-selected and spectroscopic redshift 
sample of \cl\ and spectroscopic redshift sample of \clt\, we fit a
pure de Vaucouleurs, or $n=4$ S{\'e}rsic profile, in addition to the
already existing S{\'e}rsic model.  We selected a fixed $n$ model, in
addition to the S{\'e}rsic fits, for two reasons.  First, the models
have one less degree of freedom and therefore the potential to be more
robust.  Second, a de Vaucouleurs' model has traditionally been used.
Thus, we can compare with a broader range of results from the
literature.  We plot in Figures \ref{clsizemag} and \ref{cltsizemag}
the observed size-surface brightness relation using the results from
pure de Vaucouleurs model fits.  The relation appears linear but with
a large scatter for \cl.  This scatter likely comes from a combination
of observational errors, the inclusion of non-cluster members and
systematic biases in addition to the intrinsic scatter in the
relation.  The size-surface brightness relation for \clt\ is much
tighter. The biweight scale \citep{beers90} of the scatter around the
mean relation for \clt\ is $0.42 \pm 0.05$ for the de Vaucouleurs
models and $0.36 \pm 0.03$ for the S{\'e}rsic models as opposed to
$0.76 \pm 0.10$ for the de Vaucouleurs and $0.72 \pm 0.09$ for the
S{\'e}rsic models for \cl.  We estimate the errors on the scatter
using a jackknife process \citep{beers90,lupton96}.

In general, the error bars in Figures \ref{clsizemag} and
\ref{cltsizemag} are somewhat misleading.  {\tt GALFIT} reports the
errors on the individual terms but it is well known that there are
significant anti-correlations between the errors for the surface
brightness and size \citep{hamabe87,jfk93}.  Fortunately, these errors
are almost parallel with the actual size-surface brightness relation,
so objects will be scattered along the relation.  We illustrate the
expected anti-correlation in Figures \ref{clsizemag} and
\ref{cltsizemag}.  Both figures show a good deal of incompleteness for
objects with sizes $r_h <$0\farcs1.  For \clt, this appears to be from
the magnitude limit of the sample.  However, for \cl\ we see a lack of
objects with $r_h <$0\farcs1 and faint surface brightness even though
such galaxies would be above our magnitude limit.  We assume that this
is a point spread function effect, as discussed above, and will not
use galaxies with sizes $r_h <$0\farcs1 for our analysis.

\subsection{Non-Parametric Magnitudes}
\label{petro}

The main problem with both of our current measurements of the size and
total magnitude is that they depend on fitting models to the data.  To
complement these model magnitudes, we implemented the approach of
\citet{wirth94} and \citet{wirth96}.  This approach uses the Petrosian
$\eta$ measure \citep{petrosian76} to derive a model independent
aperture.  To review, the Petrosian \(\eta = I(r)/ <I(<r)> \) computes
the ratio of the surface brightness, $I(r)$ at a particular radius,
$r$ to the average surface brightness within that radius $<I(<r)>$.
This is the inverse of the function originally developed by Petrosian,
but this version produces more tractable errors.  Given $\eta$ for our
galaxies, we need to decide on how to use it to derive an aperture.
In the literature there are two choices.  \citet{wirth94} and
\citet{wirth96} selected the aperture at which $\eta = 0.1$ to measure
the magnitude.  This aperture contains 95\% of the total light of an
exponential disk and 87\% of the light of a de Vaucouleurs profile
\citep{wirth96}.  The Sloan Digital Sky Survey
\citep[SDSS][]{strauss2002} use Petrosian derived apertures as well,
specifically the aperture containing twice the diameter at which $\eta
= 0.2$.  This aperture contains 99\% of the total light of an
exponential disk but only 82\% of the light of a de Vaucouleurs
profile \citep{strauss2002}.  Given that early-type galaxies are
dominated by de Vaucouleurs profiles, we select the approach of
\citet{wirth94} instead of the approach of the SDSS outlined in
\citet{strauss2002}.

When computing the $\eta$ distribution, we do not fit elliptical
isophotes, as did \citet{wirth96}, but use circular apertures as is
done in the SDSS \citep{strauss2002}.  \citet{lubin2001a} show that
circular apertures, when plotted against effective radii, reproduce
the results of fitting elliptical isophotes.  To compensate for the
lower signal-to-noise at larger radii, instead of using the adaptive
bin size approach of \citet{wirth96}, we choose to bin logarithmically
in radius.  Finally, to determine at what radius $\eta = 0.1$, we
interpolate our $\eta$ distributions for each object using a low order
polynomial.

This still leaves the question of correcting the aperture magnitude to
a total magnitude.  For each object, we measure the aperture where
$\eta$ crosses 0.1 and then use that size to measure a total
magnitude.  This $\eta=0.1$ magnitude is computed by interpolating
from the curve of growth of the galaxy.  This magnitude is not,
however, the total magnitude of the galaxy.  As we do not want to use
a model-dependent aperture correction, we add 0.1 mag to this
magnitude.  This would yield a magnitude 5\% larger than the true
total magnitude for an exponential disk and 3\% less than the true
total magnitude of a de Vaucouleurs profile.  As most of our galaxies
have S{\'e}rsic values of $3 \le n \le4$, this aperture correction
should, in principle, yield around 97\% to 99\% of the total light for
the galaxies.  For the rest of this paper, we will refer to this as
our Petrosian total magnitude.


\subsection{Magnitude and Size Comparisons}
\label{magcomp}

With two different models used for fitting and an additional
non-parametric total magnitude for each galaxy in each cluster, we can
independently estimate the sizes of the errors on the sizes and total
magnitudes.  The statistical errors based on the expected variance in
the photon flux for these galaxies are tiny, on the order of a
hundredth of a mag.  However, this ignores the errors caused by how
well the model describes the data or the size of aperture.  The formal
errors in the fits are also small, with a $\chi^{2}$ per degree of
freedom around $0.6 - 0.7$ for the residuals.  However, it is
straightforward to construct models that fits a given distribution
well, in the sense that the models yield a small $\chi^{2}$ per degree
of freedom, but actually deviates systematically from the model.  A
good illustration of this are galaxies which are well fit by $n=4$
models but, when fit with a S\'{e}rsic, yield a value of $n$ far from
4.  Our non-parametric measure of the total magnitude checks for this.

\begin{figure}[tbp]
\begin{center}
\includegraphics[width=3.0in]{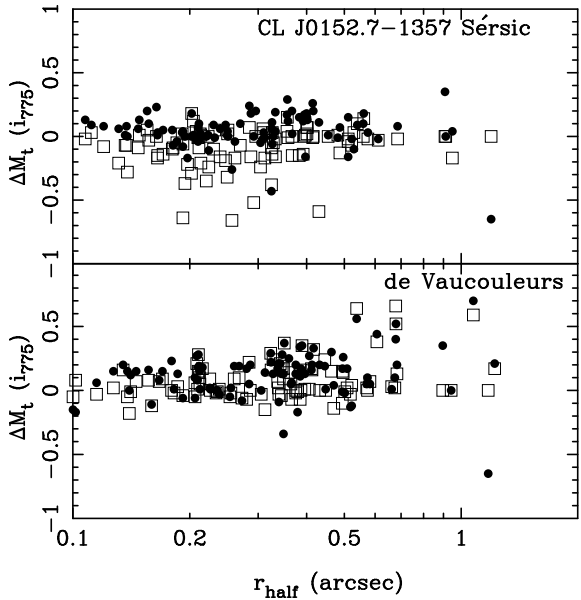}
\end{center}
 \caption[f6.eps]{The difference in the total magnitudes for
   \clt\ plotted as a function of size.  The top panel plots the
   difference between the de Vaucouleurs model magnitudes and the
   S{\'e}rsic magnitudes as squares, and the Petrosian minus
   S{\'e}rsic differences as dots.  This plot uses the S{\'e}rsic
   half-light radius as the abscissa.  In the bottom panel, we plot
   the difference between these S{\'e}rsic model magnitudes and the de
   Vaucouleurs model magnitudes, as squares, and the difference
   between the Petrosian magnitudes and the de Vaucouleurs model
   magnitudes, as dots, as a function of the de Vaucouleurs model
   half-light radius.  There are a few outliers when comparing with
   Petrosian magnitudes, see \S \ref{magcomp} for details.  Regardless
   of the magnitudes or sizes used, the overall scatter is small,
   $\simeq 0.1$ mag, with no statistically significant offset between the
   different magnitude types. }
\label{cltmagcomp}
\end{figure}

In Figure \ref{cltmagcomp}, we show the distribution of the
differences in magnitude for \clt\ as a function of the two different
sizes.  We show the difference between the total magnitude as measured
by either the S{\'e}rsic model (squares) or the Petrosian total
magnitude (circles) and the de Vaucouleurs model plotted as a function
of the half-light radius from the de Vaucouleurs model in the bottom
panel.  In the top panel, we plot the difference between the de
Vaucouleurs model magnitude and the S{\'e}rsic model magnitudes as
squares while the solid dots are the difference between the Petrosian
total magnitudes and the S{\'e}rsic model magnitudes but this time as
a function of the S{\'e}rsic model half light radius. We find 0.11
mag of scatter between the S{\'e}rsic and the de Vaucouleurs
magnitudes, 0.15 mag of scatter between the Petrosian and the
de Vaucouleurs magnitudes and only 0.06 mag of scatter in the
differences between the S{\'e}rsic and Petrosian magnitudes.  In
addition, there are systematic shifts, the de Vaucouleurs
magnitudes are on average $0.02 \pm 0.01$ mag or $0.10 \pm 0.02$ mag brighter
than the S{\'e}rsic or Petrosian magnitudes respectively.  However,
the Petrosian magnitudes are only dimmer by $0.04 \pm 0.01$ mag
than the S{\'e}rsic magnitudes.  We expect that the Petrosian total
magnitudes should be somewhat smaller than the S{\'e}rsic magnitudes
by  $\le$ 0.03 mag.

\begin{figure}[tbp]
\begin{center}
\includegraphics[width=3.0in]{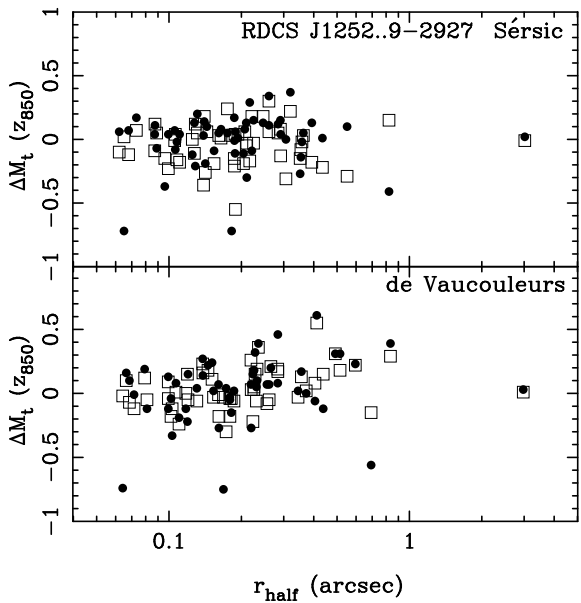}
\end{center}
\caption[f7.eps]{Same as Figure \ref{cltmagcomp} but using the
  results of \cl\ at $z=1.237$.  Compared to \clt , the
  scatter is larger, $\simeq 0.2$ mag, and there appears to be a
  small correlation between the difference in magnitude and de
  Vaucouleurs model half-light radius.  Most of this apparent
  correlation comes from galaxies with best fitting S{\'e}rsic indices
  of $n \le 2.5$.  This correlation, and the outliers, are more fully
  discussed in \S \ref{magcomp}.  }
\label{clmagcomp}
\end{figure}

We show the same quantities for \cl\ in Figure \ref{clmagcomp}.  The
scatter around the mean is larger, 0.17 mag for the S{\'e}rsic
model magnitudes and 0.21 mag for the Petrosian magnitudes as
compared with the de Vaucouleurs model magnitudes.  As before, there
are small systematic shifts.  We find that the S{\'e}rsic magnitudes
are $0.03 \pm 0.03$ mag dimmer than the de Vaucouleurs
magnitudes, while the Petrosian magnitudes are $0.07 \pm 0.03$
mag dimmer.  The scatter between the S{\'e}rsic magnitude and
the Petrosian magnitude is smaller, 0.16 mag, and the
systematic shift is also smaller, $0.04 \pm 0.02$ mag.

In Figures \ref{cltmagcomp} and \ref{clmagcomp} there is an apparent
correlation between the plotted differences and some of the measured
sizes. We examine the data  using a
Spearman Rank Correlation test.  This test yields a parameter,
referred hereafter as $t$, which is the estimate of the degree of the
correlation and can range $-1 \le t \le 1$.  Examining the data for
\cl, we find that $t = 0.30$ for the correlation between the de
Vaucouleurs magnitudes minus the S{\'e}rsic model magnitudes as a
function of the de Vaucouleurs half-light radii.  The expected error
on this measurement of the correlation, assuming that there is no true
correlation for our sample is $0.09$, yielding 3.3 $\sigma$ difference
from 0 or no correlation.  There is less of a correlation with
Petrosian magnitude, $t= 0.21 \pm 0.09$ and we find no correlation,
$t= -0.05 \pm 0.09$ between the difference in the S{\'e}rsic and the
Petrosian magnitudes as a function of the S{\'e}rsic half-light radii.
These positive correlations mean that the difference in the S{\'e}rsic
model and the de Vaucouleurs model magnitudes increases with large
size, or the S{\'e}rsic model magnitudes are fainter than the de
Vaucouleurs model magnitudes as we measure larger sizes.  This
correlation exists regardless of whether we use the S{\'e}rsic model
or de Vaucouleurs model half-light radii.  These correlations do not
appear for \clt, however.  For this cluster we find correlations with
$t=0.12 \pm 0.10$ or smaller.

\begin{figure}[tbp]
\begin{center}
\includegraphics[width=3.0in]{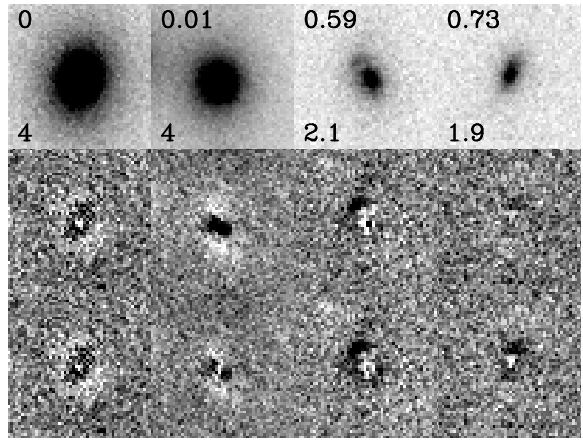}
\end{center}
\caption[stamps.eps]{Four galaxies from \clt.  For each galaxy we show
the original image (top), the residual image from the S{\'e}rsic model
fit (center) and the de Vaucouleurs model residuals (bottom).  We give
the S{\'e}rsic index in the bottom left corner of the top image, and
the $\Delta$ between the S{\'e}rsic and de Vaucouleurs model
magnitudes the top left corner.  In some cases, there are still strong
residuals after removing the S{\'e}rsic model which either point to
a higher S{\'e}rsic index being a better fit, or galaxy substructure.  }
\label{stamps}
\end{figure}

\begin{figure}[tbp]
\begin{center}
\includegraphics[width=3.0in]{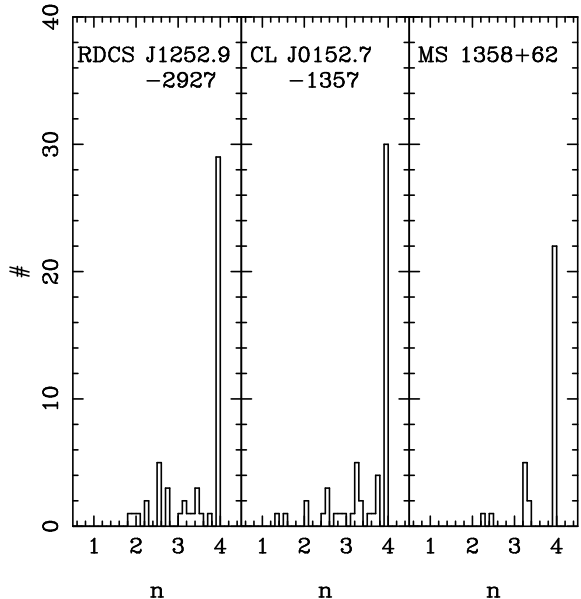}
\end{center}
\caption[f10.eps]{A histogram of the distribution of S{\'e}rsic
indices in \cl, \clt, and \ms.  There is, as expected, a strong peak
at $n=4$, as we constrain the fits to $1\le n \le 4$.  There are a
number of galaxies, however, that are classified as early-types but
have small $n$ values.  Such galaxies are the strong outliers in
Figure \ref{clmagcomp}. }
\label{nhisto}
\end{figure}

Most of the apparent correlation in Figures \ref{cltmagcomp}
\ref{clmagcomp} comes from a handful of galaxies.  In each case, the
galaxy has a large de Vaucouleurs model magnitude when compared with
the S{\'e}rsic or Petrosian magnitude.  We examined each of these in
detail.  In every case where a galaxy lies more than three standard
deviations from the mean of Figure \ref{clmagcomp}, the best fitting
S{\'e}rsic model has a low, $n \le 2.5$ S{\'e}rsic index and smaller
half-light radius.  Therefore, the de Vaucouleurs model is not a good
fit to the data.  In Figure \ref{stamps}, we show two examples of
this, along with two examples of models where the S{\'e}rsic and de
Vaucouleurs models are in good agreement.  In Figure \ref{nhisto}, we
plot the distribution of S{\'e}rsic indices for the three data sets,
\cl, \clt, and \ms.  In each case, the galaxies are identified, by
eye, as early-type galaxies.  It appears that the higher redshift
clusters have a broader range of S{\'e}rsic indices than \ms.  This
will be further discussed in \citet{postman2004}.  Nevertheless, the
de Vaucouleurs models appear to be a bad fit to a number of galaxies,
which should explain the scatter in Figure \ref{clmagcomp}.

When comparing the Petrosian magnitudes with the de Vaucouleurs model
magnitudes for \cl, there are four strong outliers.  Each of these outliers
is caused by neighboring galaxies that contaminate the Petrosian total
magnitude.  This is because our measure of the Petrosian total
magnitude assumes nothing about the underlying surface brightness
profile.  We simply use the SExtractor segmentation maps to mask out
bright pixels.  We do not attempt to compute the contribution to the
galaxy's flux by the surrounding objects.  Thus, in crowded areas
where a larger number of pixels are contaminated by luminous or
extended objects such as the brightest cluster members, our Petrosian
estimates are significantly overestimated and so appear as large
negative values in Figures \ref{cltmagcomp} and \ref{clmagcomp}.

To compliment our Petrosian, or non-parametric total magnitudes, we
can estimate a half-light radius by determining the radius where the
curve of growth contains half of the total light from the galaxy.
However, unless we make an attempt to remove the effect of the point
spread function, we will find a bias towards larger sizes as the
galaxies become intrinsically smaller.  We computed these half-light
radii and find that, on average for \cl , the Petrosian sizes are 2\%
larger than the S{\'e}rsic model sizes and 5\% larger than the de
Vaucouleurs model sizes for objects with model sizes of at least
0\farcs 15.  This size bias grows markedly at smaller model radii, as
expected.  We find no statistically significant difference between the
de Vaucouleurs model sizes and the S{\'e}rsic model sizes.  For \clt ,
we find very similar numbers.  The scatter in the size as determined
by the de Vaucouleurs model as compared with the S{\'e}rsic model is
slightly larger because of the handful of cases where the best fitting
S{\'e}rsic parameter is low, $n \le 2.5$ which causes large magnitude
discrepancies as well.  Overall, the agreement between the Petrosian
estimated half-light radii and the model radii are good.  However,
because of the bias towards larger sizes in our Petrosian half-light
radii, we rely on the model half-light radii for the rest of this
paper.

Overall, it is encouraging that our different magnitude and size
measurements yield a small scatter and little or no systematic
offsets.  Nonetheless, it is prudent not to mix the various
measurements because of the small correlations seen above.  The
Petrosian magnitudes, which yield numbers close to the S{\'e}rsic
model magnitudes, could potentially be used in conjunction with the
S{\'e}rsic half-light radii.  However, for the rest of the paper, we
will measure evolution using the sizes and surface brightnesses
determined by the model fits and we will only compare S{\'e}rsic model
fits with other S{\'e}rsic model fits, not with pure de Vaucouleurs
models.

\subsection{Rest-frame Magnitudes}
\label{restmags}

To measure the amount of evolution, we need to compare the surface
brightnesses in the same rest-frame pass band.  Most measurements of
the evolution of the FP are given in terms of the rest-frame Johnson
B, or $B_{rest}$ mass-to-light ratio.  We used the templates of
\citet{benitez2004}, which are modified versions of the templates from
\citet{cww80}, to compute the $B_{rest}$ value as a function of an
observed filter and a color.  For \cl , we computed $B_{rest} -
z_{850}$ as a function of $i_{775} - z_{850}$ while for \clt\ we
computed $B_{rest} - i_{775}$ as a function of $r_{625} - i_{775}$.
For a given galaxy in \cl\ or \clt, we then interpolate between the
colors of the templates from \citet{benitez2004} to then compute the
$B_{rest} - z_{850}$ or $B_{rest} - i_{775}$ color.  Why we chose
these particular filter combinations is illustrated in Figure
\ref{sed}, namely that the $z_{850}$ and $i_{775}$ filters match the
$B_{rest}$ filter at the redshifts of \cl\ and \clt\ respectively.
The color was selected to span, roughly, the $U-B$ color in the rest
frame of the galaxy.  This process is very similar to what was
employed by \citet{holden2004}.  In that paper, the authors found that
when the $B_{rest}$ filter was spanned by the observed colors, the
error on the magnitude transformation was on the order of 2\%.  In a
case like \cl, however, the error was higher, at around 5\%.



\section{Apparent Luminosity Evolution}

\subsection{Measuring the offset at a fixed size}

We would like to investigate how much apparent evolution there is in
the size-surface brightness relation.  First, we shall examine if the
zero-point in this relation changes with redshift.  The simplest way
to measure this is to fit a linear relation to the size-surface
brightness relation for the low redshift cluster, and then compute the
constant required to minimize the deviation around the same slope for
the higher redshift cluster.

To determine the slope of the size-surface brightness relation at low
redshift, we fit all of the data in \ms.  We then transform the \ms\
data to the $i_{775}$ or $z_{850}$ band in the manner described in \S
\ref{restmags}, and rescale the sizes as appropriate for the $z\simeq
1$ cluster, either \cl\ or \clt.  We limit both the $z\simeq 1$
cluster and \ms\ to the same magnitude.  We then apply a series of
cuts in the half-light radius, starting at a size of 0\farcs 25 and
working down to 0\farcs 1, below which the effects of the point spread
function will become important.  At each size, we compute the
zero-point in the size-surface brightness relation required to
minimize the scatter around a line defined by the slope as measured
for the whole of the \ms\ sample.  We compute this zero-point for both
the trimmed \ms\ and $z\simeq 1$ cluster samples.  We then change the
magnitude limit of the sample of \ms\ by the currently measured amount
of evolution.  We trim the data at the new magnitude limit and
continue iterating until we converge.  To compute the zero-point for
both distributions, we simply compute the biweight center
for the deviations between the average surface
brightnesses and the slope times the half-light radii.

\begin{figure}[tbp]
\begin{center}
\includegraphics[width=3.0in]{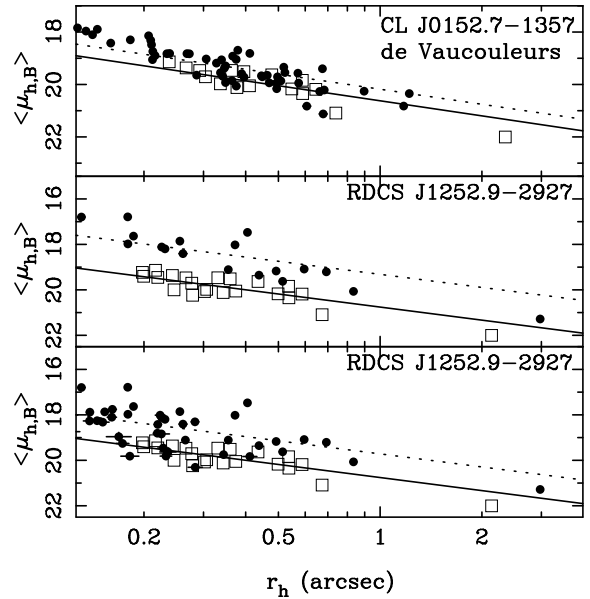}
\end{center}
\caption[f11.eps]{Size-surface brightness relation for the
  early-type galaxies in the ACS imaging data of both \cl\ and \clt\
  using de Vaucouleurs fits.  In each case, the low redshift data,
  \ms, are represented by the open squares while the filled circles
  are the data for the color selected sample from the redshift sample
  of \clt\ (top), the spectroscopic redshift sample from \cl\ (middle)
  and the color selected sample from \cl\ (bottom).  The galaxies in
  \ms\ are iteratively selected to be above the same magnitude limit
  plus the measured evolution to either \cl\ or \clt.  The solid lines
  represent the fits to the size-magnitude relation for all the data
  for \ms, while the dotted line is the relation shifted by the amount
  of measured evolution.  For \clt, we find $-0.44 \pm 0.06$ mag of
  evolution between $z=0.328$ and $z=0.837$.  We find $1.41 \pm 0.16$
  mag of evolution for the redshift selected sample between \ms\ and
  \cl\ while we find $-1.04 \pm 0.13$ mag of evolution for the
  color-selected sample.  }
\label{lum_evol}
\end{figure}

\begin{figure}[tbp]
\begin{center}
\includegraphics[width=3.0in]{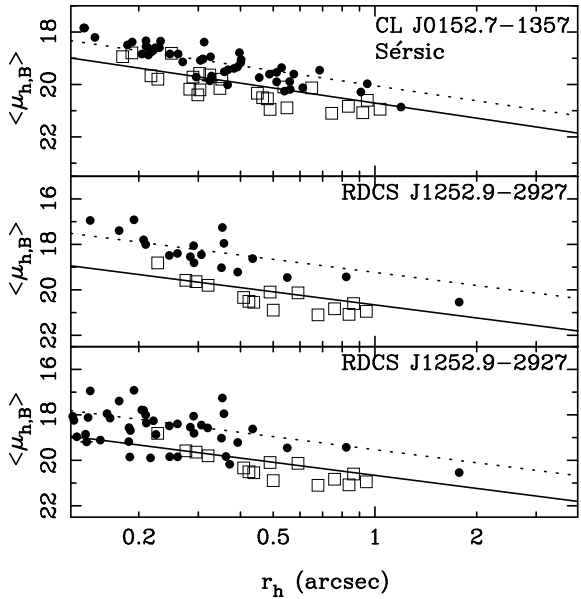}
\end{center}
\caption[f12.eps]{Size-surface brightness relation for the
  early-type galaxies in the ACS imaging data of both \cl\ and \clt\
  as compared with \ms.  The sizes and surface brightnesses are from
  the S{\'e}rsic model profiles.  See Figure \ref{lum_evol} for an
  explanation of the symbols and the lines.  The amount of measured
  evolution is similar as for the de Vaucouleurs model fits.  We find
  $-0.66 \pm 0.05$ mag for \clt\ (top) when the samples are compared with \ms,
  $-1.42 \pm 0.13$ mag of evolution for the redshift-selected
  sample of \cl\ (middle), and $-1.08 \pm 0.13$ mag of evolution for the
  color-selected sample of \cl\ (bottom).}
\label{lum_evol_s}
\end{figure}

We find $-1.04 \pm 0.13$ mag of evolution is required to match the
size-surface brightness relation of \ms\ when comparing the de
Vaucouleurs model results for galaxies with $r_h > $0\farcs 15 and
$z_{850} < 24.5$ in \cl.  In rest-frame $B$, $z_{850} = 24.5$
corresponds to $M_B = -20.0$ at $z=1.24$ and $M_B = -18.8$ at $z=0$
after assuming the FP evolution of \citet{vandokkum2003}.  We plot, in
Figure \ref{lum_evol}, the data from \ms\ as open squares, both for
the pure de Vaucouleurs models.  The results for \cl\ are represented
by solid dots.  The solid line represents the fit to the whole data
for \ms.  The dotted line is the same but shifted by $-1.04 \pm 0.13$
$B$ mag.  This shift was calculated by minimizing the scatter around a
line with the slope specified by the whole sample.  Repeating the
process but using the S{\'e}rsic model profiles, we find $-1.08 \pm
0.13$ $B$ mag of evolution (see Figure \ref{lum_evol_s}).  We note
here, that, as we discussed above, \citet{kelson2000a} allows a larger
range of the parameter $n$ in their fits.  All galaxies with $n >4$ in
S{\'e}rsic models in \citet{kelson2000a} were replaced by the
corresponding de Vaucouleurs model.

We make the same computation for \clt\ and find $-0.44 \pm 0.06$
mag of $B$ band evolution for $r_h > $0\farcs 15, the size
limit for the smallest galaxies in \ms , and $-0.66 \pm 0.05$
mag of evolution in the $B$ band for the S{\'e}rsic models.  If
we restrict ourselves to only those galaxies in \cl\ with redshift
information and with $z_{850} < 23$, we find $-1.41 \pm 0.16$
mag (de Vaucouleurs model) and $-1.42 \pm 0.13$ (S{\'e}rsic
model) mag of $B$ band evolution for $r_h > $0\farcs 15. 

We summarize the results of these fits in Table \ref{tevol}.  In that
table, we also show the results for a selection of cutoff radii and
for each of the samples of \cl.  We selected a number of half-light
radii thresholds.  We decided to not examine galaxies with $r_h <
$0\farcs 10 as they are close in size to the full-width at
half-maximum of the point spread function and the samples appear
incomplete at those sizes, see \S \ref{model}.  However, there are few
galaxies in the sample of \ms\ that have $r_h < $0\farcs 15, so we
selected that for our practical limit.  We included a cutoff of $r_h >
$0\farcs 20 along with $r_h > $0\farcs 10 to illustrate that our
results are insensitive to the exact threshold selected.

The error we quote is the error on the mean shift between the low
redshift size-surface brightness relation and the high redshift
relation.  We assume no error on the slope measurement, simply taking
the scatter around the mean relation and dividing it by the square
root of the number of galaxies in our high redshift sample.  This
raises the question of how our particular choice of the slope of the
size-surface brightness relation affects our results.  It appears,
when looking at Figures \ref{lum_evol} and \ref{lum_evol_s}, that the
slope of the size-surface brightness relation is different for the
different samples.  We compare the data from \ms\ and \cl\ using the
slope of the best fitting relation for \cl\ instead the slope of the
relation for \ms.  We find almost the same results with small changes,
$\simeq 0.01$ mag, in the measured evolution.  For \clt , our changes
in measured evolution are larger, on the order of $0.03$ mag to, at
most, $0.05$ mag.  This provides an additional source of uncertainty
which we will add in quadrature.

\subsection{Comparison with the Fundamental Plane}

\citet{vandokkum2003} measured the velocity dispersions for three
galaxies in \\ CL~0848+4453, a low mass cluster of galaxies at
$z=1.276$.  \citet{holden2005} measured velocity dispersions for four
galaxies in \cl, which, when combined with \citet{vandokkum2003},
yields a sample of seven objects with $\bar{z} = 1.25$.  
Using those values along with sizes and surface brightness from HST
imaging, the authors measured the offset in the FP from $z=0$ to the
redshift of the cluster.  The evolution corresponds to a change in the
mass-to-light ratio (M/L) with redshift of $-0.98\, z$ in the rest-frame
$B$ filter, or $(-1.06 \pm 0.06)\, z$ mag of $B$ band evolution
at a fixed mass.  For \cl , this should correspond to $-1.31$ 
mag of evolution in the rest-frame $B$ band.  To compare our
relative measurements between clusters at different redshifts with
those made using the FP, we need to compare our measured evolution
with the Coma cluster, the $z=0$ baseline for \citet{vandokkum2003}
and \citet{holden2005}.

To measure how much evolution occurs between \ms\ and Coma, we
implemented the same procedure as we used for comparing \cl\ and \clt\
with \ms.  For the data for the Coma cluster, we use the $B$
half-light radii and surface brightnesses, listed in Table A 7 of
\citet{jfk96}.  In \citet{kelson2000a}, the authors compare the
techniques used to measure the half-light radii and surface
brightnesses in \citet{jfk96} with the two-dimensional fitting
techniques they used and find no significant offsets.  Thus, we
compare the Coma radii and surface brightnesses directly with the
values tabulated in \citet{kelson2000c}.  We measure a shift of $-0.35
\pm 0.07$ mag in the rest-frame $B$ band between Coma and \ms.
These measurements were done with the whole sample of Coma and \ms, as
both are magnitude limited, redshift selected samples.

The amount of evolution expected between $z=0$ and $z=0.328$ is $-0.35
\pm 0.06$ mag in rest-frame $B$ using the best fitting relation
from \citet{vandokkum2003}, in good agreement with the $-0.31 \pm
0.12$ mag expected from \citet{vandokkum96} and the $-0.33 \pm
0.12$ mag from \citet{kelson97}.  Our result of $-0.35 \pm
0.07$ mag is in excellent agreement but with smaller
statistical error bars. The systematic errors are the same as the
FP, see \citet{kelson2000a}.

\begin{figure}[tbp]
\begin{center}
\includegraphics[width=3.0in]{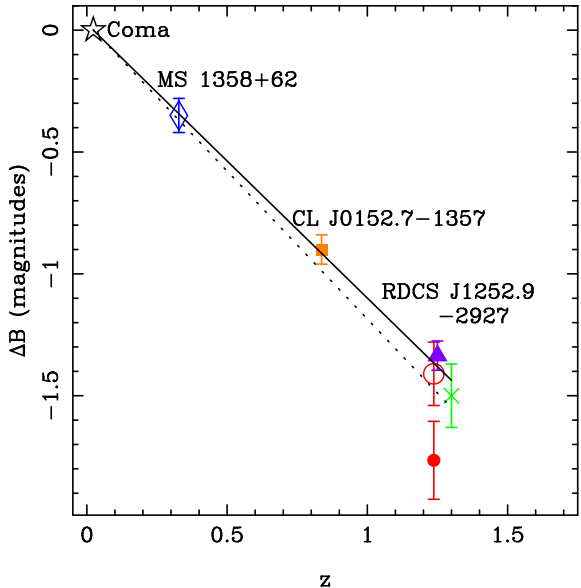}
\end{center}
\caption[f13.eps]{Amount of $B$ luminosity evolution from the change
  in the zero-point of the size-surface brightness relation with
  redshift.  The results for \ms, represented as an open blue diamond
  at $z=0.33$, shows the measured amount of evolution with respect to
  Coma using our size-surface brightness measurements.  The orange
  square represents the average of the S\'{e}rsic and de Vaucouleurs
  results for the $r_h \le $0\farcs 15 sample of \clt.  The solid red
  circle is average for the spectroscopically-selected sample of \cl\
  and the open red circle represents average for the color-selected
  sample. We plot, with a solid line, a fit to the evolution of \ms,
  \clt, and the color-selected sample of \cl.  The dotted line shows
  the fit to the the same data except using the
  spectroscopically-selected sample of \cl.  For both lines, the error
  bars at $z=1$ represents the error in the slope.  The results for
  either fit agree with the FP measurements in CL~0848+4453
  \citep{vandokkum2003}, shown with a green cross, and \cl\
  \citep{holden2005}, shown with a purple triangle.}
\label{ml_evol}
\end{figure}

We add the measured evolution between Coma and \ms\ to the measured
evolution we find for \cl\ and \clt.  We plot both in Figure
\ref{ml_evol} along with the FP results from \citet{vandokkum2003} and
\citet{holden2005}.  We fit a line to the data for \clt\ and for the
color-selected sample of \cl\ and find $(-1.13 \pm 0.15) z$ mag of $B$
evolution, very close to the $(-1.06 \pm 0.06) z$ mag found with the
FP.  We find $(-1.42 \pm 0.33) z$ mag of evolution if we use the
redshift-selected sample of \cl, a result not statistically different
from the FP because of the much error estimate.

Overall, our agreement with the measurements from the FP are good,
with the caveat that it is clear that sample selection plays a
critical role.  Our first sample of early-type galaxies in \cl\ was
constructed by finding all galaxies with the correct morphologies that
also fall within the color-magnitude sequence of \cl.  Restricting our
sample to galaxies with sizes of $r_h >$0\farcs 15, we find a result
in almost perfect agreement with restuls from the Fundamental Plane
\citep{vandokkum2003,holden2005}, regardless of whether we use
S{\'e}rsic or de Vaucouleurs models.  If, however, we restrict
ourselves to those galaxies identified only by redshift, we find more
evolution at the highest redshifts.  The redshift selection was made
using a $K$ magnitude-limited sample, with some color selection.  As
shown in Figures \ref{clcm} and \ref{clsizemag}, almost all early-type
galaxies with $z < 23$ are included.  Therefore, it is puzzling why
such a sample would appear to show so much more evolution.

\subsection{Scatter in the Size-Surface Brightness Relation}
\label{scatter}

\begin{figure}[tbp]
\begin{center}
\includegraphics[width=3.0in]{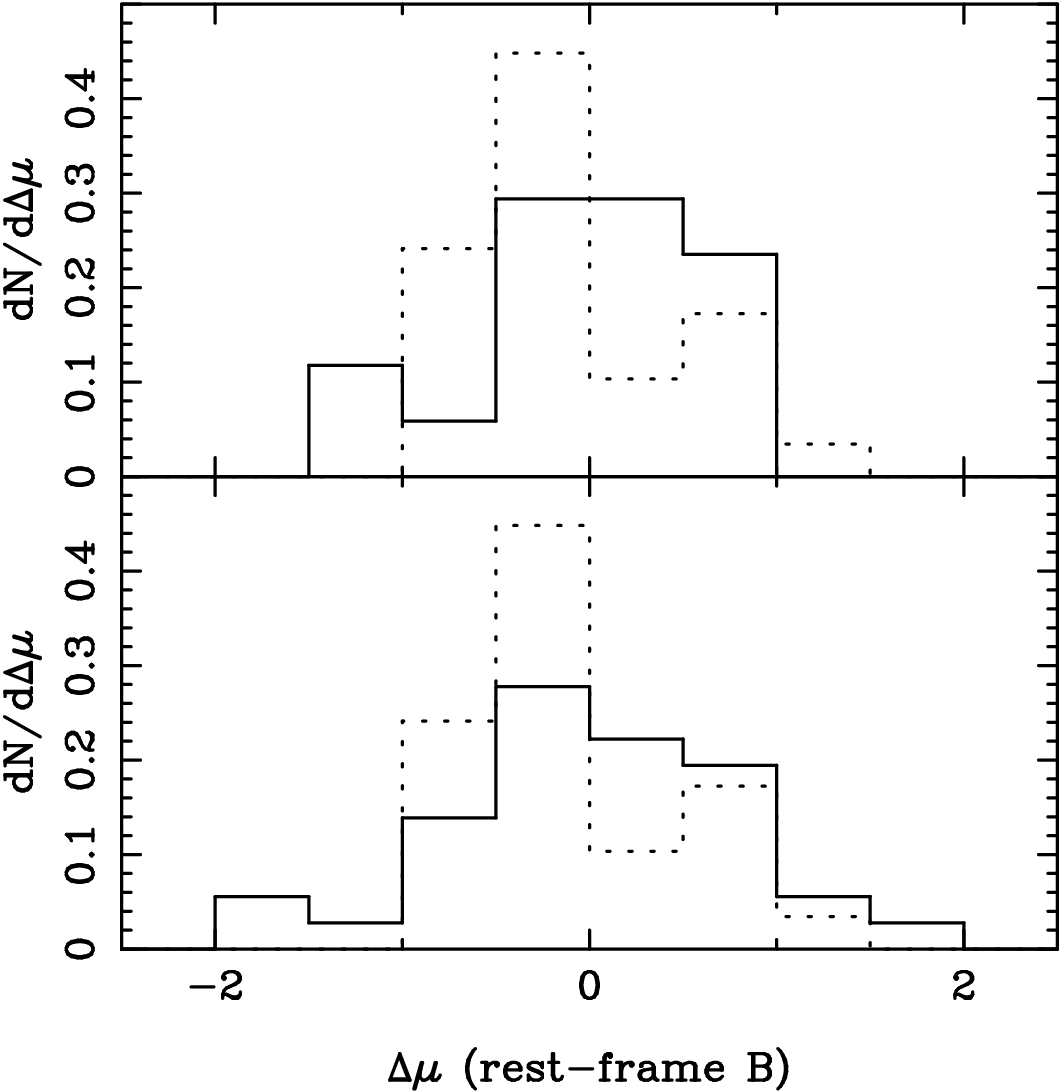}
\end{center}
\caption[f14.eps]{The deviations from the mean size-surface
  brightness relation for \cl\ (solid line) and \ms\ (dotted line)
  plotted as normalized histograms.  All galaxies in \cl\ with $r_h
  >$0\farcs 15 are included.  The top panel contains all galaxies in
  the redshift sample and the bottom panel contains all those in the
  color selected sample, all galaxies within the dotted lines in
  Figure \ref{clcm}.  The scatter for \cl\ is $0.76 \pm 0.10$ mag for
  the color selected sample and $0.65 \pm 0.12$ mag for the redshift
  selected sample, in contrast to the $0.57 \pm 0.07$ mag found for
  \ms. }
\label{fitdevs}
\end{figure}

In Figure \ref{fitdevs}, we plot the the histogram of the scatter
around the size-surface brightness relation.  This is the difference,
in magnitudes, between the measured surface brightnesses and those
predicted by the size-surface brightness relation.  We plot the
residuals for \cl\ as compared with those from \ms, using the de
Vaucouleurs model sizes and surface brightnesses.  The solid line
represents the results for \cl, while the dotted line is scatter for
the comparison sample of \ms.  We plot the histogram containing only
confirmed cluster members in the top panel while the bottom panel
shows all color selected galaxies.  In both panels, the histograms are
normalized to one.  For this figure, we plot all galaxies with
projected sizes of $r_h > 0$\farcs 15 at $z=1.237$.  The magnitudes
for the galaxies in \ms\ have been brightened by the measured $-1.04
\pm 0.13$ mag of evolution in the $B$ band for the bottom panel
and $-1.44 \pm 0.15$ mag in the top panel.  

\begin{figure}[tbp]
\begin{center}
\includegraphics[width=3.0in]{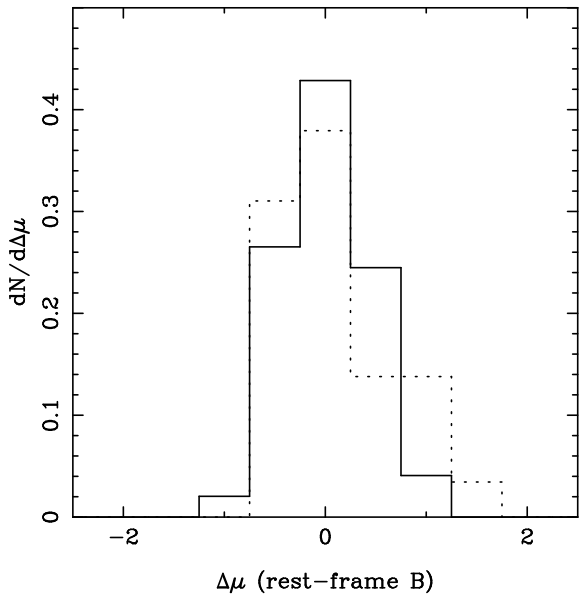}
\end{center}
\caption[f15.eps]{The deviations around the mean size-surface
  brightness relation for \clt\ (solid line) and \ms\ (dotted line).
  This Figure is similar to Figure \ref{fitdevs}.  All galaxies with
  redshifts in \clt, solid points in Figure \ref{cltcm}, and $r_h
  >$0\farcs 15 are included.  The scatter for \clt, $0.42 \pm 0.05$
  mag, is smaller than the results for \ms.}
\label{fitdevs2}
\end{figure}

Examining Figure \ref{clsizemag} again, however, we see that the
objects below the mean relation have, on average, a larger error in
both the size and average surface brightness than those above the mean
relation.  Thus, a low tail towards large deviations is expected.  At
brighter magnitudes there are three galaxies brighter than expected
based on the distribution of \ms.  The observed scatter is, as
mentioned before, $0.76 \pm 0.10$ mag around a fixed size.  Given that
the errors in our total magnitude measurements are, based on the
comparisons between the de Vaucouleurs model magnitudes and the
Petrosian total magnitudes, around $0.2$ mag, most of the scatter must
be intrinsic or systematic.  For \clt, there are 55 early-type
galaxies cluster members that meet our magnitude and size criteria.
In Figure \ref{fitdevs2}, we plot the distribution of the two sets of
galaxies, and note that the scatter is $0.42 \pm 0.05$ mag.  For
comparison, we find in the sample from \ms a scatter of $0.57 \pm
0.07$ mag.

We calculate for each of our scatter measurements an error based on
the jackknife of the data as recommended by \citet{beers90}.  It seems
that the scatter measurements are statistically very different.  For
example, the difference between the scatter in \clt\ and in \ms\ is
$0.15 \pm 0.06$ mag, a 2.5 standard deviation difference.  If we make
the assumption that scatter in the size-surface brightness relation
comes from a Gaussian parent population, we can use the F-ratio test
to compare the variances of two different scatter measurements.  When
we compare the scatter for \cl, $0.75 \pm 0.10$ mag, with that of \ms,
$0.57 \pm 0.07$ mag, we find that, despite the $0.19$ mag difference,
the odds of the two populations being drawn from the same parent
population is around 2\%, not a statistically significant difference.
We find a better agreement when comparing \clt\ with \ms, a 15\%
probability that variance measurements of the two samples are drawn
from the same distribution, despite the apparent 2.5 standard
deviation difference.

In \S \ref{acs}, we discussed the accuracy of our morphological
classifications.  At fainter magnitudes, we found that our scatter in
classification was higher, as measured by the disagreement amongst the
classifiers.  The inclusion of later-type galaxies could cause a bias
in the estimate of the evolution.  We measured the evolution using
only those galaxies with unanimous agreement amongst the four
classifiers and found a larger result, $-1.46 \pm 0.13$ mag for the
color selected sample.  This agrees well with the $-1.44 \pm 0.15$ mag
found for the redshift selected sample.  The inclusion of late-type
galaxies should bias towards more evolution, as spiral structure and
large disks usually hold star formation, yet the inclusion of the more
uncertain classifications actually lessens the amount of evolution.
However, the lower surface brightness galaxies are also the galaxies
with more uncertainty in their classification.  Thus, by removing the
galaxies with higher uncertainy in their classification, we naturally
remove the lower surface brightness galaxies.  The scatter for the
color-selected sample, with uncertain objects removed, is $0.64 \pm
0.12$ mag, very close to $0.65 \pm 0.12$ mag we find for the
spectroscopically selected sample.  Thus, we raise the possiblity that
the difference in the two samples could come from classification
errors as the majority of uncertain classifications are in the $23 <
z_{850} < 24.5$ mag range, and below the cutoff for the
spectroscopically selected sample.  Removing these less certain
classifications lower the scatter and raises the amount of evolution
measured in the color-selected sample.  However, these objects have,
on average, a lower surface brightness, 0.35 mag in $z_{850}$, than
the objects with unanimous classifications.  These lower surface
brightnesses could also explain why the classifications are uncertain.

\section{Discussion}

We observe, using the size-surface brightness relation, $(-1.13 \pm
0.15) z$ $B$ mag of luminosity evolution from $z=0$ to $z=1.237$.  If
we restrict our highest redshift sample to only those galaxies with
spectroscopic redshifts, instead of the large sample of galaxies with
red colors, we find $(-1.42 \pm 0.33) z$ mag of evolution, which is in
statistical agreement with our other result.  If we use only the color
selection for \cl, we find that both of the measurements agree well
with the FP results at the same redshifts
\citep{vandokkum1998,vandokkum2003,holden2005} and, thus, with the
overall trend in luminosity evolution found from other size-surface
brightness studies \citep{schade97,schade99,lubin2001c}. The overall
agreement points towards the conclusion that the majority of massive
elliptical galaxies at these redshifts are passively evolving, in
agreement with their color evolution.

Given that we find slow luminosity evolution for the majority of
cluster early-type galaxies in our samples, this is another indication
that the luminosity-weighted redshift of formation is high for these
objects.  \citet{vandokkum2003} finds $z_{f} > 2.5$, though for a
different cosmology, using the FP, while \citet{holden2005} finds $z_f
2.3^{+0.3}_{-0.2}$ for certain assumptions about early-type galaxy
evolution.  As the evolution found in that paper agrees with the
evolution in offset of the size-surface brightness relation, we would
find a similar result.  This epoch of formation is also in agreement
with \citet{blakeslee2003} from the colors of galaxies in \cl.  In
general, the large scatter on the size-surface brightness relation
prevents us from making significantly more accurate measurements of
$z_{f}$.  Since we selected galaxies for red colors in \cl, we could
very well be missing an important part of the cluster early-type
population, namely those galaxies that will become red early-types but
are bluer than our selection criteria at the epoch of the
observations.  This would bias our results to higher redshifts.

The redshift selected sample of \cl, however, shows much more
evolution, $(-1.42 \pm 0.33) z$ mag as compared with the
color-selected sample.  It is possible that this is a selection
effect.  Namely, it is easier to measure redshifts for higher surface
brightnesses objects.  We attempted to mitigate this by trimming the
redshift-selected sample at $z_{850} < 23$, where the
redshift-selected sample is almost complete, containing 17 out of 20
early-type galaxies with $r_h>$0\farcs 15, regardless of color.  One
possible explanation for this result is the scatter around the mean
size-surface brightness relation.  

We observe that the apparent scatter for \cl\ around the size-surface
brightness relation is larger than for \clt.  This result is not
statistically siginificant though an increase in the scatter would be
expected at higher redshifts.  There are a couple of lines of evidence
that, at $z\simeq1$, there are elliptical galaxies that show evidence
of recent star formation.  Such objects would show a smaller than
average $M/L$.  Both \citet{vandokkum2003} and \citet{tran2003} find
galaxies morphologically classified as early-type galaxies but with
strong Balmer absorption lines indicating recent starbursts.  The
example galaxy in the Fundamental Plane sample of
\citet{vandokkum2003} has a much lower than expected $M/L$ value, a
factor of $\simeq$13 lower than the massive galaxies in that sample,
and \citet{holden2005} finds a large scatter in the $M/L$ value for a
purely luminosity selected sample, much larger than the scatter for
lower redshift clusters such as Coma or \ms.  The inclusion of a
handful of such galaxies in our sample would increase the scatter but
disentangling them requires spectra or good information.
Spectroscopic evidence for recent bursts of star-formation in \cl,
which could explain both the larger scatter and the larger apparent
evolution when using only the spectroscopically-selected sample, will
be discussed in a future paper \citep{rosati2005}.

\section{Summary}

We have investigated the size-surface brightness relation for two
clusters of galaxies, \cl\ at $z=1.237$ and \clt\ at $z=0.837$.  For
each cluster, we identified, by eye, a sample of early-type galaxies
across a broad range in color and in magnitude \citep{postman2004}.
Using our ACS imaging, which covers the rest-frame $B$, we created a
magnitude limited sample for each cluster.  The magnitude limits
correspond to approximately $0.3 L^{\star}$ for \cl\ and $0.5
L^{\star}$ for \clt.  For \clt, we selected galaxies based on the 102
spectroscopic redshifts \citep{demarco2003PhDT}.  \cl\ has 36
spectroscopic members \citep{demarco2003PhDT,rosati2004,rosati2005}
and we expanded the sample for that cluster using a color selection.
Our color criteria for \cl\ utilizes the color-magnitude relation and
scatter from \citet{blakeslee2003}, including all galaxies within two
standard deviations of the mean relation.  In each case, we chose the
observed magnitude limit in the filter that corresponds most closely
to the rest-frame $B$ band.

We fit both S{\'e}rsic and de Vaucouleurs models to each galaxy in our
two cluster samples.  These models provide total magnitudes and
half-light radii with the point spread function broadening removed.
We converted the total magnitudes to the average surface brightness
within the half-light radius.  In addition to the model total
magnitudes, we implemented a third total magnitude measurement.  We
used the Petrosian magnitudes as implemented in \citet{wirth96} and
found excellent agreement between these non-parametric measurements
and the S{\'e}rsic model magnitudes.  The disagreement with the de
Vaucouleurs models in general involved a small systematic of $0.04 -
0.1$ mag depending on the cluster.  In addition, we found that
the difference between the de Vaucouleurs model magnitudes and either
the S{\'e}rsic or Petrosian magnitudes correlated with the model sizes
for one of the two clusters.  In general, we find that our overall
trends are robust, regardless of which size and magnitude we used for
determining the surface brightness as long as we compare a given
cluster measurement with a measurement made using the same model
in a different cluster.

We measured the amount of rest-frame $B$ magnitude evolution by
computing the shift in apparent surface brightness at a fixed size.
We compare the clusters in this paper to the results from \ms\ in
\citet{kelson2000a}, which contains both S{\'e}rsic and de Vaucouleurs
average surface brightnesses and half-light radii.  For \ms, we
compute the amount of evolution required to match the size-surface
brightness relation between \ms\ and either \cl\ or \clt, with the
data trimmed at a fixed size and magnitude limit.  We apply this
correction to the data in \ms, recompute which galaxies would fall
above our magnitude limit, and repeat until we converge.  We also
apply this to \ms\ and the $B$ band Coma cluster data of
\citet{jfk96}.  For that, we measure $-0.35 \pm 0.07$ mag of
evolution, in excellent agreement with the Fundamental Plane, or FP,
measurements \citep{vandokkum96,kelson97,kelson2000c}.  The rest of
the evolution measurements, summarized in Table \ref{tevol}, show
results consistent with the FP in most cases.  We find that the
average surface brightness at a fixed sized changes with redshift as
$(1.13 \pm 0.15) z$, very close to the $(-1.15 \pm 0.13) z$ found with
the FP.  Only when we examine the subset of galaxies in \cl\ selected
by spectroscopic membership do we find a deviation.

The overall trend in the luminosity evolution that we find agrees with
this.  At the redshift of \cl, we are likely within a $2-3$ Gyrs of
observing epoch when most of the stars formed in cluster early-type
galaxies.

We would like to thank Greg Wirth for advice on measuring Petrosian
magnitudes and for supplying copy of his software package.  BH would
also like to thank S. Adam Stanford and Daniel Kelson for useful
discussions.  ACS was developed under NASA contract NAS5-32865, this
research was supported by NASA grant NAG5-7697.  We are grateful to
K.~Anderson, J.~McCann, S.~Busching, A.~Framarini, S.~Barkhouser, and
T.~Allen for their invaluable contributions to the ACS project at JHU.


\begin{deluxetable}{lrrrrr}
\tablecolumns{6}
\tablecaption{Summary of Measured Evolution from \ms \label{tevol}}
\tablehead{
\colhead{Cluster} & \colhead{z} & \colhead{Membership\tablenotemark{a}} &
\colhead{Min. $r_h$\tablenotemark{a}} & \colhead{$\Delta \mu\, (n=4)$\tablenotemark{c}}
& \colhead{$\Delta \mu\, (1 \le n \le 4) $\tablenotemark{c}} \\
\colhead{} & \colhead{} & \colhead{} &
\colhead{} & \colhead{($B_{rest}$ mag)} & \colhead{($B_{rest}$ mag)} \\
 }
\startdata
\clt & 0.837 & z   &  0\farcs 20  & -0.42 $\pm$ 0.06  & -0.67 $\pm$ 0.06 \\
\clt & 0.837 & z   &  0\farcs 15  & -0.44 $\pm$ 0.06  & -0.66 $\pm$ 0.06 \\
\clt & 0.837 & z   &  0\farcs 10  & -0.57 $\pm$ 0.06  & -0.68 $\pm$ 0.05 \\
\cl & 1.237 & color & 0\farcs 20  & -0.97 $\pm$ 0.14  & -1.17 $\pm$ 0.13 \\
\cl & 1.237 & color & 0\farcs 15  & -1.04 $\pm$ 0.13  & -1.08 $\pm$ 0.13 \\
\cl & 1.237 & color & 0\farcs 10  & -1.09 $\pm$ 0.11  & -1.06 $\pm$ 0.11 \\
\cl & 1.237 & z    &  0\farcs 20  & -1.32 $\pm$ 0.16  & -1.28 $\pm$ 0.11 \\
\cl & 1.237 & z    &  0\farcs 15  & -1.41 $\pm$ 0.16  & -1.42 $\pm$ 0.13 \\
\cl & 1.237 & z    &  0\farcs 10  & -1.53 $\pm$ 0.16  & -1.47 $\pm$ 0.13 \\
\enddata
\tablenotetext{a}{This column notes whether the galaxies in the
  sample were selected based on spectroscopic redshifts (z) or based
  on the galaxies' colors.}
\tablenotetext{b}{Minimum half-light radius for inclusion in the
  sample.}
\tablenotetext{c}{Average change in surface brightness measured for
  the sample based on de Vaucouleurs models ($n=4$) or S\'{e}rsic
  models ($1 \le n \le 4$.)}
\end{deluxetable}

\end{document}